# Ferrielectricity in an Organic Ferroelectric

J. F. Scott,[1] F. D. Morrison,[1] Rebecca Clulow,[1] Philip Lightfoot,[1] Aurora Gherson,[1] Sylvia C. Capelli,[2] Michael R. Probert,[3] S. Sahoo,[4,5] J. S. Young,[4] R. S. Katiyar,[4]

+

[1]Schools of Chemistry and Physics, St. Andrews Univ., St. Andrews, Scotland KY16 9ST2

[2]Rutherford Appleton Lab, ISIS Neutron & Muon Source, Didcot OX11 0QX, Oxon, UK

[3]Newcastle Univ., School of Chemistry, Newcastle Upon Tyne NE1 7RU, UK

[4]Speclab, Dept. Physics, University of Puerto Rico, San Juan, PR 90036 USA

[5]Present address: Institute of Physics, Bubaneshwar, Orissa, India

Abstract

We report ferrielectricity in a single-phase crystal, TSCC -- tris-sarcosine calcium chloride [$(CH_3NHCH_2COOH)_3.CaCl_2$]. Ferrielectricity is well known in smectic liquid crystals but almost unknown in true crystalline solids. Pulvari reported it in 1960 in mixtures of ferroelectrics and antiferroelectrics, but only at high fields. TSCC exhibits a second-order displacive phase transition near $T_c$ = 130 K that can be lowered to a Quantum Critical Point at zero Kelvin via Br- or I-substitution, and phases predicted to be antiferroelectric at high pressure and low temperatures. Unusually, the size of the primitive unit cell does not increase. We measure hysteresis loops and polarization below T = 64 K and clear Raman evidence for this transition, as well of another transition near 47-50 K. X-ray and neutron studies below Tc = 130K show there is an antiferroelectric displacement out of plane of two sarcosine groups; but these are antiparallel displacements are of different magnitude, leading to a bias voltage that grows with decreasing T. A monoclinic subgroup $C_2$ may be possible at the lowest temperatures (T<64K or T<48K), but no direct evidence exists for a crystal class lower than orthorhombic.

I.     Introduction

Ferroelectrics and antiferroelectrics are well known among low-symmetry crystals and are directly analogous to ferromagnets and antiferromagnets, with polarizations P replacing magnetizations M. However, ferrielectrics, with antiparallel dipole ordering but net polarizations are very rare, except among liquid crystals. TSCC (tris-sarcosine calcium chloride) has a second-order displacive paraelectric (PE) to ferroelectric (FE) phase transition near $T_c$ = 130 K.[1,2] Early work by Bornarel and Schmidt[3] showed that tris-sarcosine calcium chloride also has a structural phase transition at ca. 700 MPa hydrostatic pressure at T=293 K and that the phase boundary extrapolates to << 40K at atmospheric pressure (their studies were limited to T > 77 K). They suggested that the high-pressure phase is antiferroelectric (AFE) but offered no evidence for that. Some data on the hypothesized PE, AFE, FE triple point near T = 177 K and P = 503 MPa were provided, but the structures and full phase diagrams remain enigmatic. Those authors pointed out that an AFE structure in TSCC need not involve a doubling (or other multiplication) of the primitive unit cell, since the paraelectric and ferroelectric phases already have 4 formula groups (and 12 sarcosine molecules) per unit cell. Thus, an AFE distortion might realign the polarizations to be antiparallel (say, 2 up and 2 down) in contrast to the 4 parallel dipoles per ferroelectric unit cell. Such an unusual structure also could permit simultaneous FE and AFE ordering, either a ferrielectric canted arrangement (as in ferrimagnets[4]), or a Lieb-Mattis type antiferroic, un-canted linear polarization geometry with, for example, 3 polarizations up and 1 down. This would help explain the polarization data P(T) with strong built-in potential (bias field) reported by Fujimoto et al.[5] Although several authors have suggested that the high-pressure ambient temperature phase of TSCC is the same antiferroelectric structure as the ambient-pressure low-temperature phase, that is impossible: We emphasize that Fujimoto et al. found that P remains nonzero down to very low T, ruling out a centric $C_{2h}$ symmetry, but found that the phase at high pressures and ambient temperatures is probably $C_{2h}$ and definitely non-polar.

Our earlier Raman studies[6,7] revealed no additional vibrational modes in TSCC at T = 4 K compared with T = 300 K. This supported the idea that any low-T structure(s) had no increase in unit cell size. However, we found unambiguous evidence by other techniques for the so-called AFE transition at T = 64 K, as well as additional anomalies near T = 47 K, but no proof that these anomalies are phase transitions in either a crystallographic or thermodynamic sense.

The aim of the present report is to provide new electrical and Raman data that confirm the transition at 1 atmosphere pressure and T ca. 64-70 K. Additional Raman and pyroelectric anomalies verify the transition near 50 K. These transitions are important, because in bromine- or iodine-substituted TSCC, replacement of Cl-ions lowers $T_c$ to zero Kelvin (ca. 80% Br), and the quantum critical point dynamics are not quantitatively compatible with a conventional uniaxial ferroelectric; hence, the structure (AFE or other) below 47 K is pertinent, and in particular, one should clarify whether it remains polar.

The general idea here is that a ferrielectric crystal would offer a new kind of switching device, with single-polarity switching (both positive or both negative) and high threshold voltages.

II.     Earlier Experiments and Inconsistencies

1. Low temperature and high pressure (Roth et al.)[8]

Roth et al. showed several things about the Raman spectra at low temperatures and high pressures: Firstly, they emphasized that the high-pressure phase had very nearly the same number of Raman lines as did the ferroelectric $Pn2_1a$ phase or the paraelectric Pnma phase, and therefore that it is probably an antiferroelectric non-polar $P2_1/a$ phase ($C_{2h}$) with the same Z=4 formula groups (12 sarcosines) per primitive cell. In our opinion this is at best a *non-sequitur* and very likely incorrect. In fact, although the ferroelectric and so-called antiferroelectric phases would have nearly the same number of vibrational modes, the supposed $C_{2h}$ phase would have many fewer *Raman-active* modes. This arises for two reasons: The vibrations in $C_{2h}$ have parity, and the odd-parity modes are not Raman-allowed, which would decrease the number of Raman lines by approximately half. And in addition, many of the modes in the ferroelectric phase will be split into transverse-longitudinal doublets (TO/LO), giving rise to an even larger difference between the number of modes expected in the $Pn2_1a$ and $P2_1/a$ phases. So rather than confirming the predicted symmetry from Schmidt, the work of Rost et al. actually contradicts it. The paradox here is that usually a FE/AFE transition doubles the unit cell and hence approximately doubles the number of Raman lines; but in this particular case, the predicted $C_{2h}$ AFE structure would approximately halve the number of Raman lines. Hence it probably can be ruled out, despite other evidence supporting it. The second important point is that no one has previously observed any antiferroelectric double hysteresis loops P(E) in TSCC in any phase. Definitions of antiferroelectricity generally require this.

The only argument given by Roth et al. for monoclinic $C_{2h}$ symmetry is that they found only two clear symmetry varieties of vibrational modes, whereas they argue polar structures should have four; however, this argument is generally false – in the present case $C_{2h}$ and $D_2$ would each have four different irreducible representations, of which two would be Raman-inactive in $C_{2h}$ but possibly extremely weak in a $D_2$ structure only slightly distorted from ferroelectric $C_{2v}$, and the subgroups $C_2$, $C_s$, or $C_i$ each have only two. In any event, the low-T phase is strongly polar and hence cannot be $C_{2h}$.

Schmidt has published an argument[9,10] claiming that the lack of double hysteresis loops is expected because the transition from PE to AFE or FE to AFE is first-order. The collective view of the present authors is that the presence or absence of double loops in the hysteresis of an AFE material has nothing to do with the order of a transition to another phase from the AFE state.

2. 64K transition

There is now ample evidence for a structural phase transition at 64 K (about 73 K in TSCC with a few percent Br). Although the Raman spectra below 64 K are indeed very similar to those at high pressure, the phase boundary inferred by Bornarel and Schmidt and by Schmidt does not extrapolate to 64 K but to a much lower temperature.

It is worth noting that the soft mode in the ferroelectric phase abruptly vanishes at the high-pressure phase transition. This confirms that the transition is first-order, but it raises symmetry questions about the point group of the high-pressure phase. In the ferroelectric phase the soft mode is of totally symmetric symmetry; this is always a requirement for any non-reconstructive group-subgroup transition ("Worlock's Lemma"). If the high-pressure phase is a subgroup of the ferroelectric phase, the soft mode must remain Raman-active. The fact that it is not seen implies either that it has become diffusive (order-disorder), or that the high-pressure phase has a higher symmetry than the low-pressure FE phase. Note that doubling of the primitive cell does not affect this argument.

3. Raman symmetries

Roth et al. argue that they see only two kinds of Raman spectra, thus ruling out a $P2_12_12_1$ piezoelectric structure (which would have four) and supporting a $C_{2h}$ $P2_1/a$ structure at low temperatures. This disagrees with our neutron scattering and polarization results, which show that the structure at low T is polar. In addition this is still another *non-sequitur*, since the correct symmetry argument does not reveal intensity ratios for different polarizabilities, merely which ones are nonzero. A transition from orthorhombic-orthorhombic is *not* excluded by the data of Roth et al.

To summarize previous work: A series of conjectures and *non sequiturs* has unfortunately led various authors to label both the room-temperature high-pressure phase and the low-temperature atmospheric-pressure phases of TSCC as antiferroelectric $C_{2h}$. This has been inconsistent with both the number and symmetry of Raman lines and the lack of double-loop hysteresis, and with the switching at low temperatures.

4. New Raman work:

Raman data as a function of temperature for a TSCC single crystal with 3% Br (chosen to match most recent work) are shown in Figs.1a and 1b. In Fig. 1a we see a strong decrease in intensity of the lowest-frequency modes near 65 cm$^{-1}$ and increase in their splitting, and in addition a dramatic disappearance of the mode near 40 cm$^{-1}$ at 76 K, together with the appearance of an intense new mode near 115 cm$^{-1}$. This new mode at ca. 115 cm$^{-1}$ is an indication of the high-pressure phase, conspicuous in the data of Roth et al. It is a "signature mode" that lends some support (but not enough – see below) to the idea that the phase at P = 1 atm and T<64K is the same as that at P>700 MPa and T=293K. However, it does imply that these two structures are similar.

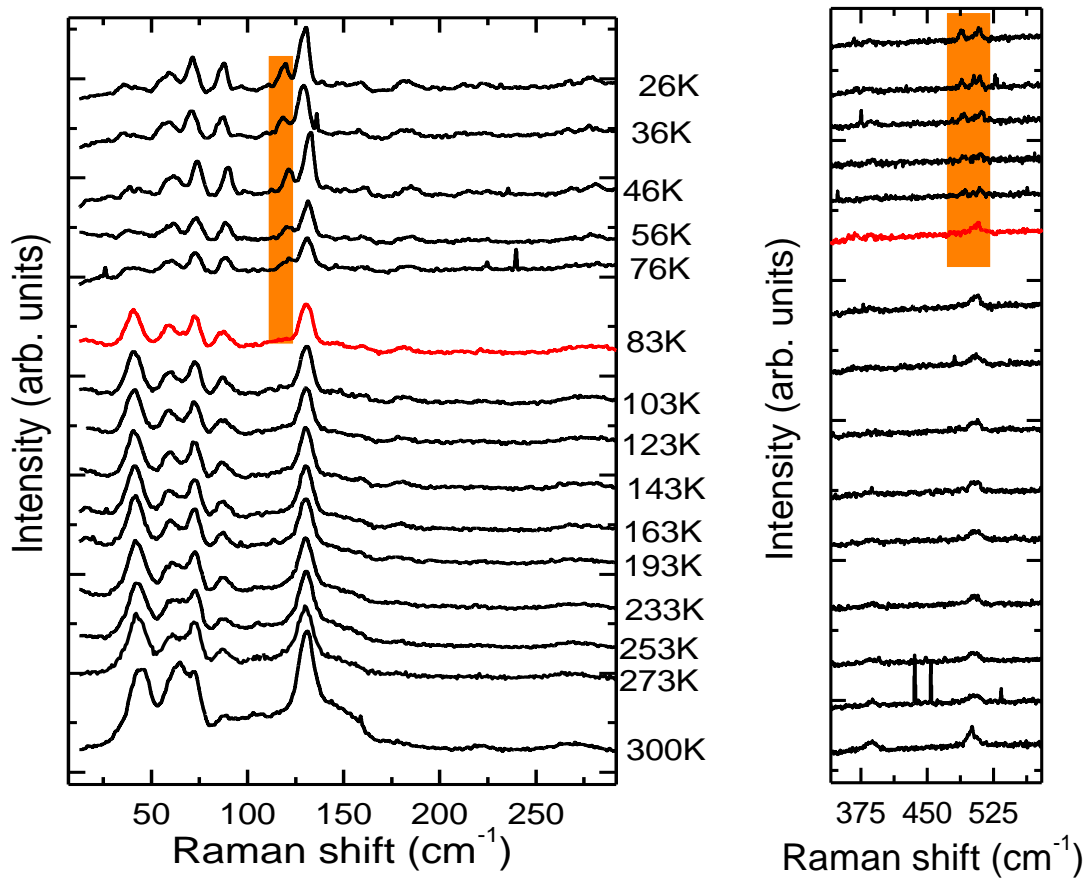

Fig. 1: (a) Left side. Raman data (intensity versus energy) in TSCC with 3% Br at various temperatures, showing anomalies in intensity and energy for low frequency vibrations near 40 and 115 cm$^{-1}$ at ca. T = 75 K. (b) Right side. Raman data (intensity versus energy) in TSCC with 3% Br at various temperatures, showing anomalies near 500 cm$^{-1}$ at ca. T= 50K.

In Fig.1b (same temperature runs) we also see that a mode near 500 cm$^{-1}$ splits into a doublet ca. 25 cm$^{-1}$ apart below ca. T = 70 K.

We interpret the disappearance of the vibrational mode at 40 cm$^{-1}$ and the appearance of a new mode near 115 cm$^{-1}$ as clear signatures of a phase transition near 70 K.  n.b.,  The splitting of the mode near 500 cm$^{-1}$ at about 47 K confirms the lower phase transition.  Neither of these phase transitions is likely to involve unit cell doubling, which would roughly double the entire number of Raman lines.  They suggest instead a small reduction in symmetry, from say orthorhombic $C_{2v}$ to monoclinic $C_2$. However, there are no degenerate vibrations in orthorhombic crystals; therefore the small splitting in the mode near 500 cm$^{-1}$ is not a removal of degeneracy, but could arise from changes in transverse-longitudinal (TO/LO) mode splittings.

III.    Electrical measurements

1. Switching polarization loops P(E)

In an effort to demonstrate antiferroelectric double-loop hysteresis, we subjected TSCC single-crystal samples up to 4.0 kV across 0.3-1.4 mm (E = 33 kV/cm) at T = 40 K. This is 10x the known coercive field of 3 kV/cm for TSCC.  Results are shown in Fig. 2ab.

If the phase at high P or low T is $C_{2h}$, what should be seen at such fields is ferroelectric single loops between 130 K and 64 K and double loops or no loops at all below 64 K.  On the other hand, if the low-T (or high-P) phase is $P2_12_12_1$ or a lower antiferroelectric polar phase, one should see both FE single loops and AFE double loops.  It is of course possible that the structure is ferrielectric[11] with canted polarizations or different numbers of dipoles up and down (say 3 up and 1 down per unit cell). Ferrielectricity is rare but has been known since the early work of Cross[11] and Pulvari.[12] Most recently it has been reported in multiferroic $DyMn_2O_5$[13,14] and in $CuInP_2S_6$.[15] Ferrielectric transitions are also well known in ferroelectric liquid crystals.  Gleeson's group has studied the study of a four-layer smectic that can be ferro-, ferri-, or antiferro-electric.[16,17] The four layers can be ordered 4-up, 3-up/1-down, or 2-up/2-down, possibly in analogy with the four polarizations per unit cell in TSCC. She emphasizes that the response of these systems is highly asymmetric with regard to voltage polarity. Similarly, Gou and Rondinelli have considered[18] the case of an artificial ABABAB... superlattice in which the two kinds of layers can have FE, FI, or AFE ordering.

Based upon the loops observed in Fig.2, and the absence of any P(E) loops along the other crystallographic axes (up to 33 kV/cm along the c-axis), we conclude that the phase below 64 K is ferroelectric and not antiferroelectric (it could be both simultaneously with canted polarizations P); but the Raman data and polarization data prove that it is not $C_{2h}$.

This was established previously in very nice work by Fujimoto et al.[4] but ignored in subsequent work by Roth et al.[8] The $D_2$ point group $P2_12_12_1$ is not possible, because it is not polar, but lower symmetries are.

Note especially that the loop is not symmetric about V=0; instead an extremely large bias field is observed (built-in potential) such that the entire loop is shifted ca. 10 MV/m to the +V side. We confirmed that this is an asymmetry between the (0 1 0) and (0 -1 0) axes (and not an artefact) by mechanically inverting the sample. The data in Fig.2 resemble those in an antiferroelectric, but with only one polarity switching. We suggest that this is not an accident, and that the correct description of this system is probably "ferrielectric," with two polarization components of the same magnitude, one of which switches and one of which does not because of quite different values of coercive field $E_c$.

2. Four-State Model Hypothesis:

TSCC has a complex structure with two inequivalent sarcosine sites, often designated S1 and S2. In analogy with the 4-state model of polarizations in smectic liquid crystal ferroelectrics,[16,17] we consider as a speculation a model for low-temperature TSCC that permits the four formula groups per unit cell (12 sarcosines) at the sites usually labelled S1 and S2 to order in pairs at different temperatures. (See also Gou and Rondinelli.[18])

As with the model of Gleeson's group, this would give four different states: At T > 130 K the Pnma paraelectric phase has no polarizations ordered. As T is lowered below 130K, 64K, and 48K, additional sarcosines might order. However, this would predict polarization steps of integer ratios, and we do not see such steps. Therefore any analogy with Gleeson's liquid crystal system seems only qualitative.

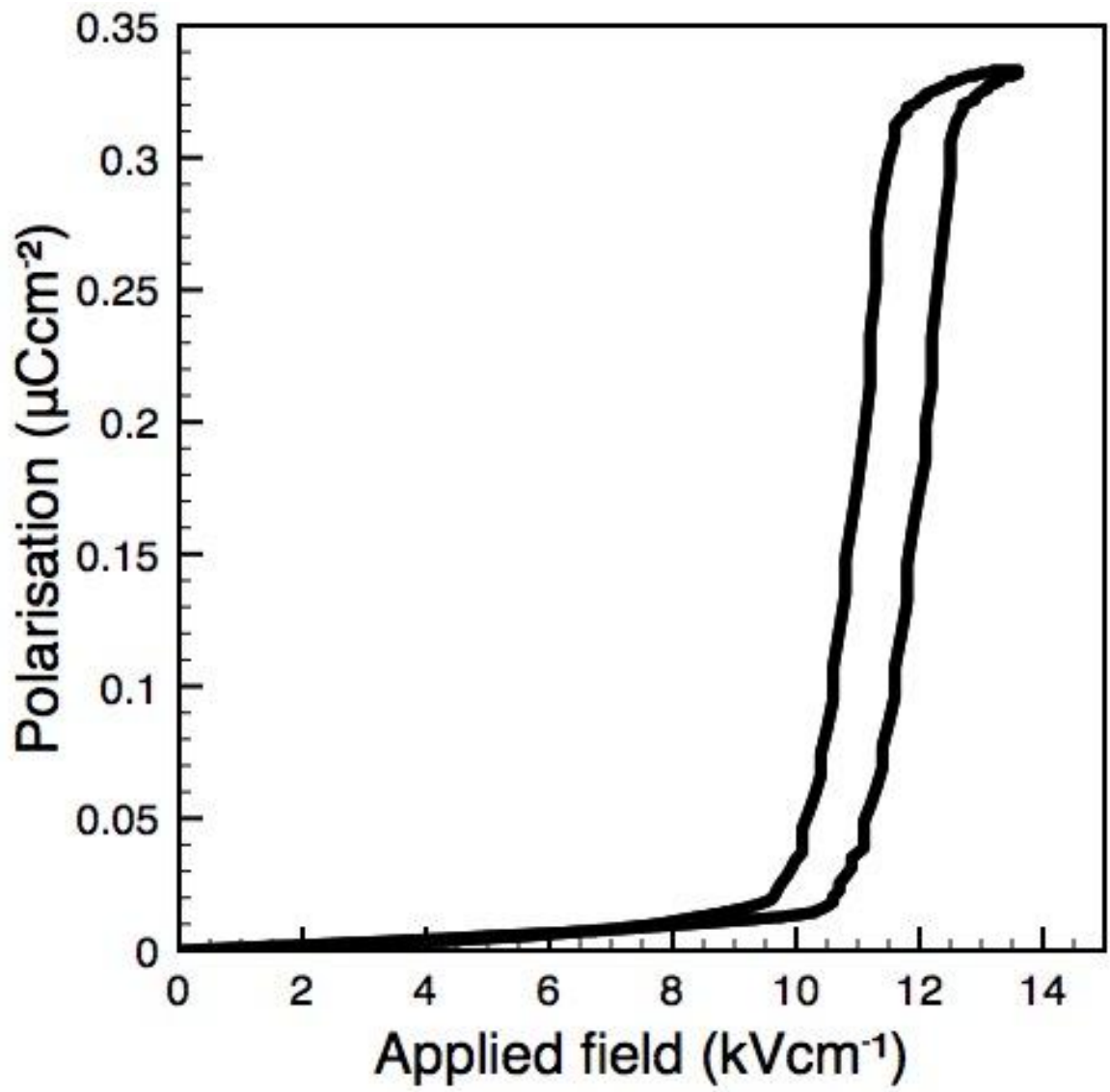

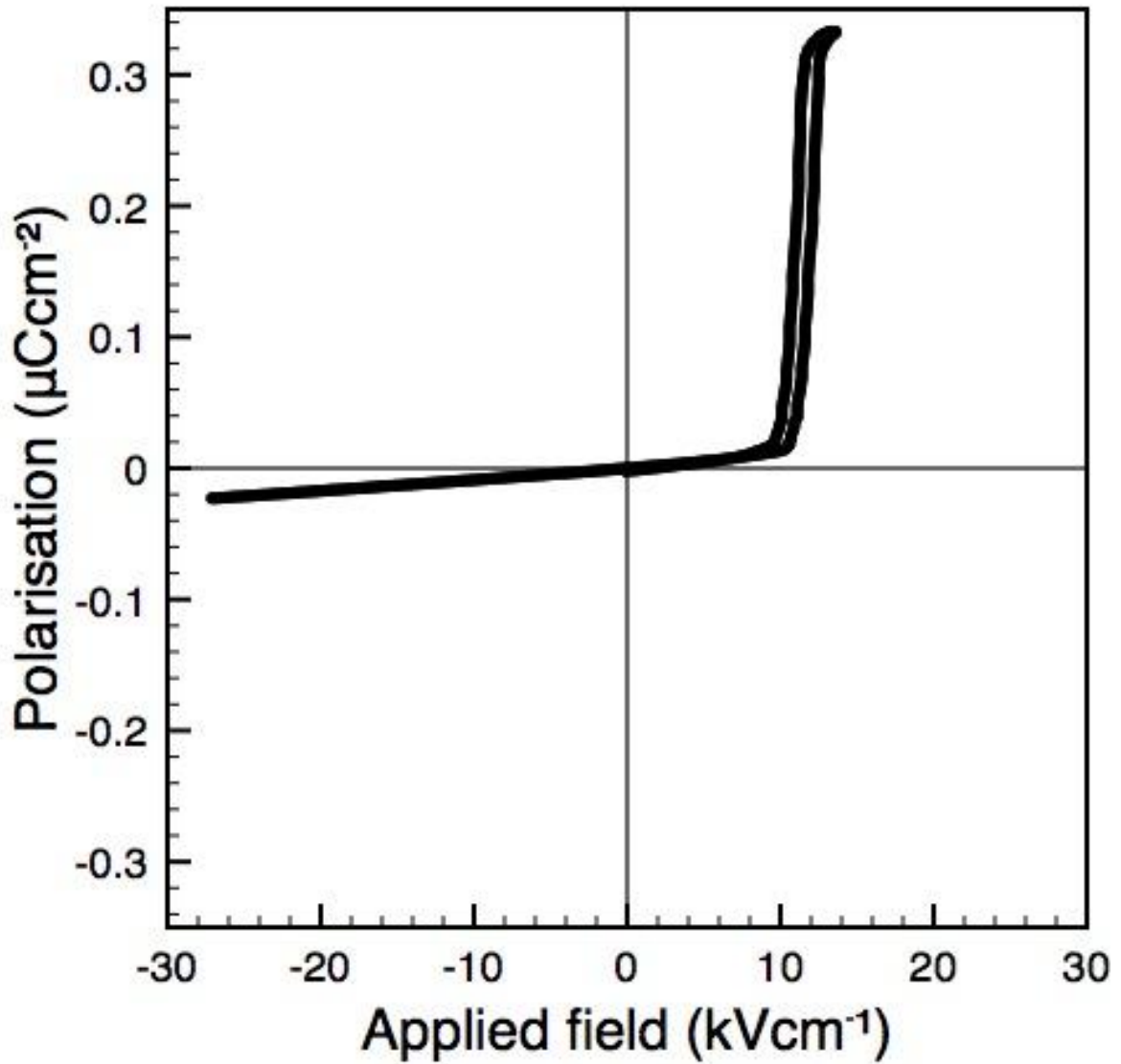

Fig. 2: (a) Upper figure. Polarization data versus field P(E) for TSCC along the polar +b-axis; E = 30 kV/cm along [0 1 0] with 4 kV across 1.4 mm at T = 40 K and 100 Hz. Switching is observed in the form of a single antiferroelectric-like loop (i.e. a ferroelectric loop with a very large bias field) from ca. 8-12 MV/cm, in agreement with ref.4. However, this occurs only for positive voltages. (b) Lower figure. E = 30 kV/cm along the –b axis [0 -1 0]. If we invert the sample mechanically, switching is observed for only negative voltages. For fields along the pseudo-hexagonal c-axis no hysteresis is observed up to 100 kV/cm.

We can do this only by physically rotating the sample (Fig.3b).

This model is compatible with the original discoveries of the transition near 64 K, first by Haga et al.[19] and subsequently by Lee et al.;[20] the former was via specific heat and yielded an entropy change of 1.16 J/mol.K (about half that of 2.51 J/mol.K at $T_c = 130$ K), but incorrectly inferred that this low-T phase is the same as the high-pressure phase, whereas the latter showed that it was an effect of proton dynamics and NOT a static structural phase transition. Regrettably that study did not present data below T = 50 K (an unusual stopping point), an unfortunate coincidence since we found[2,6] some indication o another transition near 48 K.

3. Other AFE-like transitions with only one loop

Very recently Randall's group at Penn State have also found AFE-like loops that occur for only one voltage polarity in the famous AFE sodium niobate, "which in fact is an antiferroelectric structurally, but as it switches under the forward electric coercive field, remains in a metastable ferroelectric state and avoids the switch back to the antiferroelectric, thereby missing the tell-tale features of the double hysteresis loop."[21] Sodium niobate is one of the two materials originally described by Pulvari as ferroelectric. In calcium zirconate/sodium niobate they observe[22,23] a related smooth FE-AFE transition with increasing Ca/Zn %.

4. Pyroelectric measurements

If we invoke a polar structure at temperatures below 64 K, it is useful to measure the absolute polarization and show that it is nonzero. This can be done via second harmonic generation or by pyroelectric measurements.

The latter are summarized in Fig.3. Here we measure P(T) at T = 77 K as 0.3 $\mu C/cm^2$, in agreement with earlier work. Below ca. 64 K the polarization does not vanish (ruling out any centric structure); instead it increases an additional 10% as temperature lowers, in agreement with ref. 4.

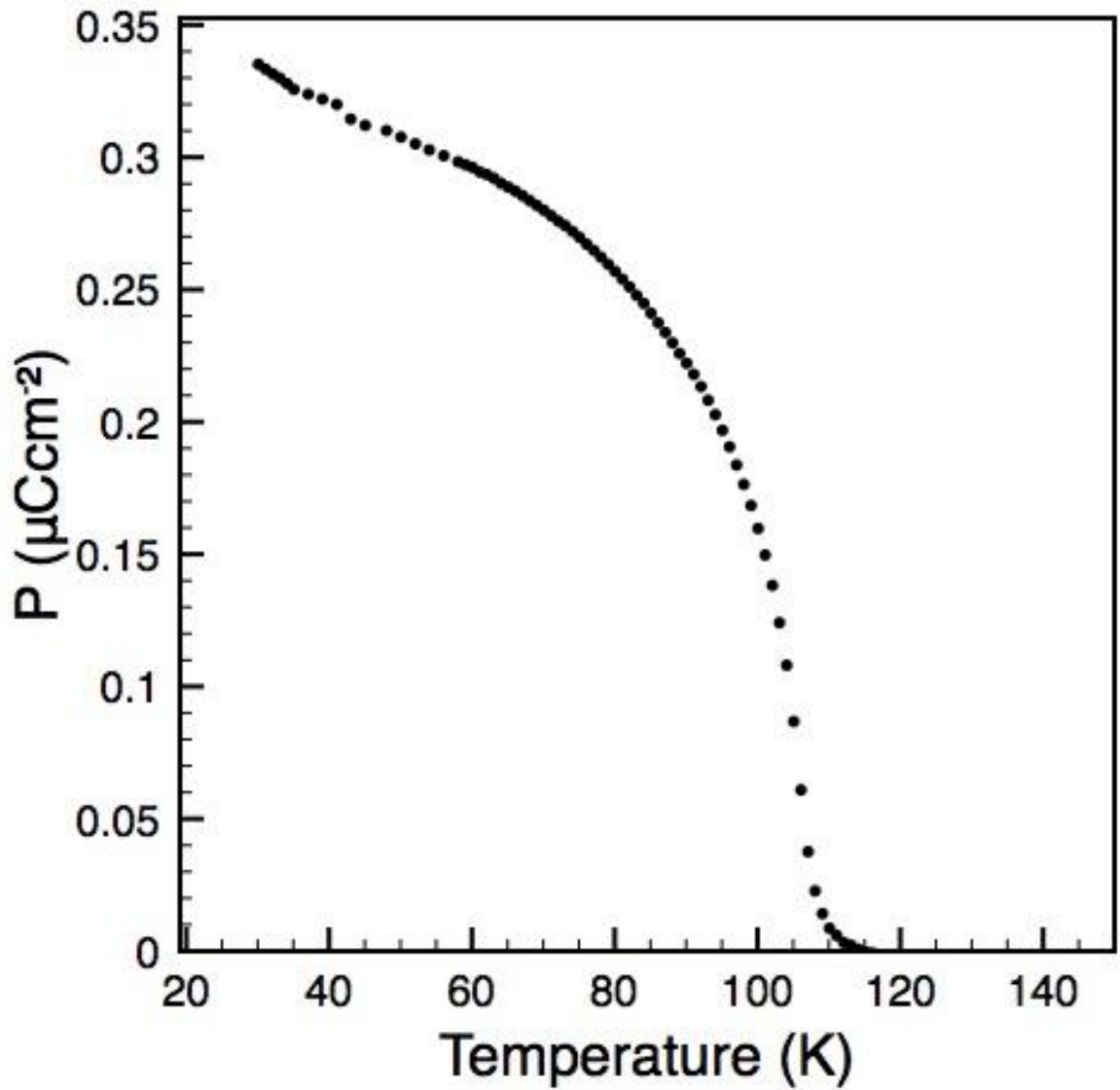

Fig.3. Pyroelectric response P(T) at zero voltage along the polar b-axis. Note that the maximum polarization P measured this way of 0.33 µC/cm² agrees with that from the polarization switching loops in Fig.2 and that the lowest temperature phases are pyroelectric, and in fact that P increases in these phases..

5. PUND data

Whenever ferroelectric polarization hysteresis data are unusual it is wise to record PUND

data. PUND is an acronym describing separate measurements of the four parts of a switching loop, first employed by Ramtron Corp.:[24] P – Positive non-switching voltage; U – Up switching voltage from –P to +P; N – Negative non-switching voltage; D – Down switching voltage. It reliably discriminates between true ferroelectric polarization reversal [which is actually detected as a charge, by integrating displacement current i(t)] and artifacts such as real injected charge [also detected by integrating leakage current J(t)]. Our PUND data are shown in Fig.4. Here we employ a triangular voltage drive, with intentional separation of 5 ms between the +V and –V pulses. We conclude that there are no artifacts.

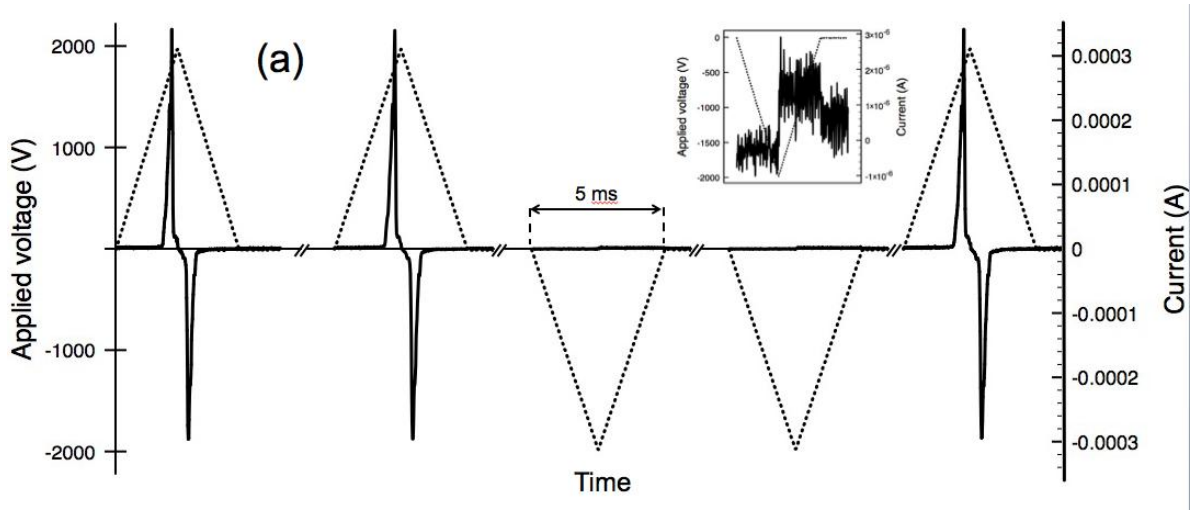

Fig. 4. PUND data on TSCC at T=77K.

IV. The transition at 47 K

An additional phase transition near 47K at atmospheric pressure was first inferred from dielectric and thermal data by the present group.[2,6] However, in retrospect, if we go back and scrutinize the early specific heat data of Haga et al.,[19] it seems likely that the divergence near 64 K has a complexity on its low-temperature side, perhaps indicative of another phase transition with weak entropy change ($< 0.5$ J/mol); see Fig. 5.

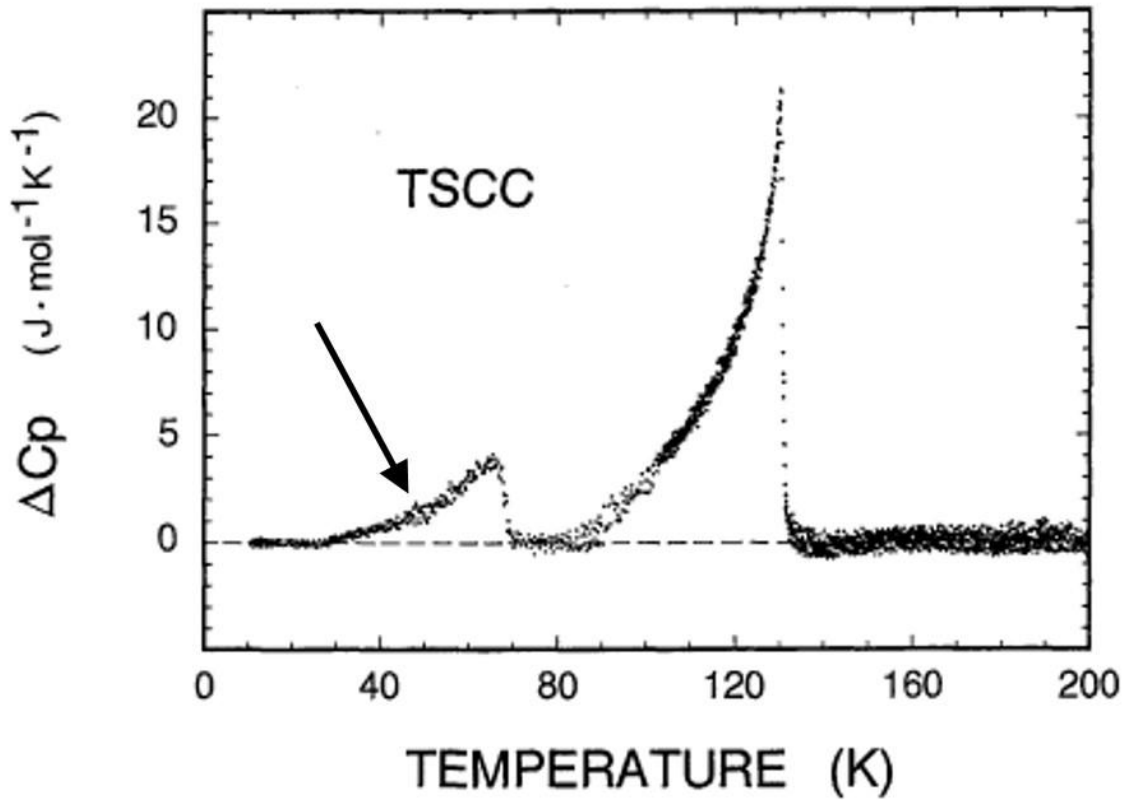

Fig. 5. Specific heat data near 64K and 130K in TSCC, modified from Haga.[19]

V. Structural determination

1. X-rays

With all the conflicting evidence listed above, the obvious test was to do structural determination on single-crystal samples of pure TSCC (undoped). These are summarized below. The single-crystal XRD was intended to determine crystal class (orthorhombic, monoclinic, or triclinic) by precisely measuring lattice angles to see if one, two, or all three were exactly 90 degrees. No symmetry lowering from orthorhombic was observed via X-ray, and no evidence for a phase transition at 64K or 48K.

2. Neutron structural data: Figure 6-8.

The neutron scattering data were intended to determine the space group symmetries below T=48K and for 48K < T < 64K. It had been suggested that perhaps only H-ions move, and since the entropy change at 64K is significant (Fig.5), an order-disorder transition was suspected. The fact that the entropy change at 64K is nearly half that at 130K may not be a coincidence and may relate to the two different sarcosine sites. However, no phase transition is observed, only a continuous smooth increase in rotation angle of ca. 3 degrees.

Fig. 6 shows that the two sarcosines (2a, 2b) on the mirror plane above $T_c$ do not move in opposite antiferroelectric directions, as might be expected in an AFE transition. Instead, one (2a) moves almost not at all, whereas the second (2b) moves in a way to compensate the off-plane displacement of sarcosine 1, which is already out of plane above $T_c$. However, they move by slightly different amounts y(T), adding to a net (small) polarization along the polar axis. This is probably the origin of the large P(V) bias field observed in Figs.2ab; the bias field $E_b(T)$ scales with this net displacement y(T). However, since we do not know the effective charge of the displaced ions, we cannot yet confirm that this small net displacement gives the measured quantitative $E_b(T)$.

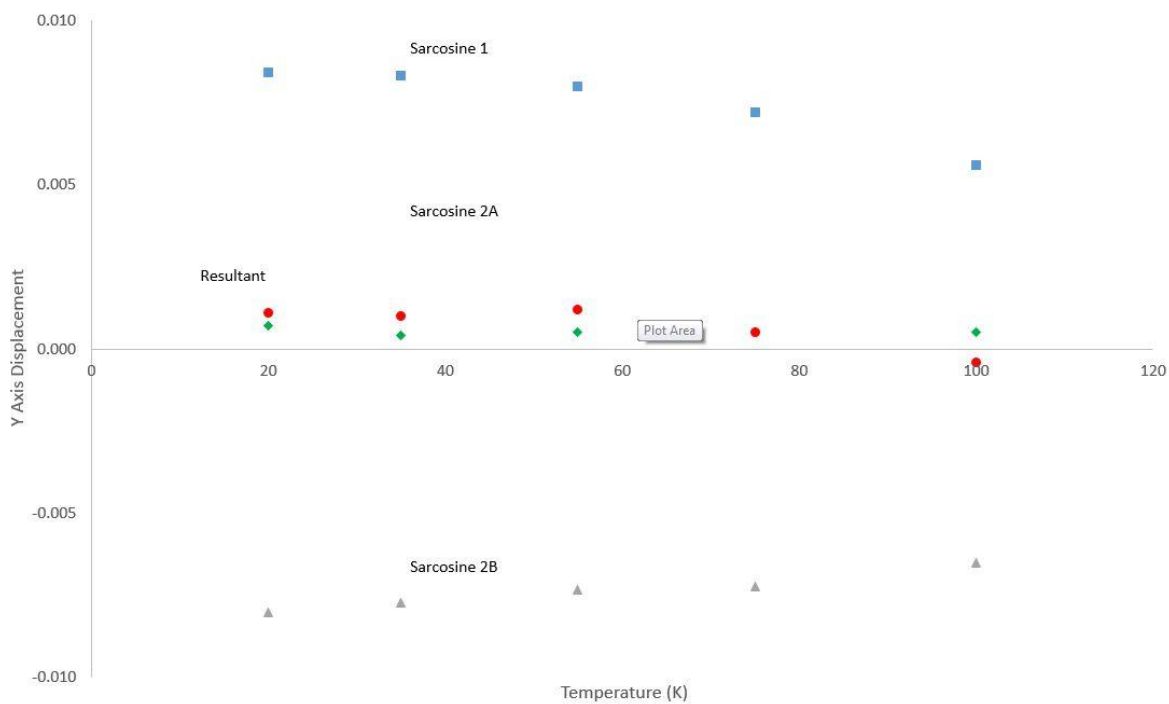

Fig. 6. Sarcosine displacements below Tc = ca. 120K; vertical axis units are % of the b-axis lattice constant; horizontal scale is temperature (K) just underneath the middle (y=0) axis.

Fig. 7 shows the resulting TSCC structure in the ferroelectric phase.

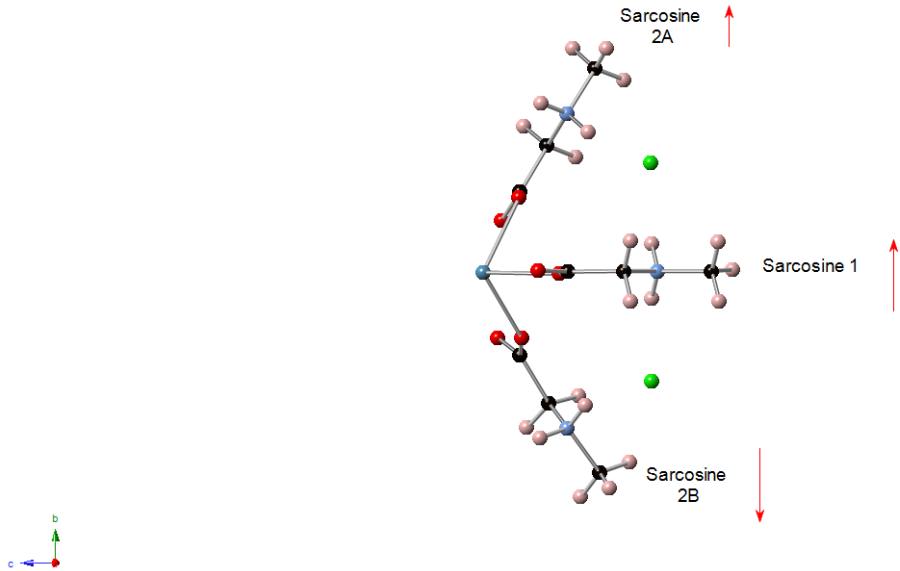

Fig. 7. Resultant TSCC structure below Tc = ca. 120K. Although there is a small net ferroelectric polarization, as shown in Fig. 6 the main displacements are out of phase (antiferroelectric). Fig. 8 shows this more clearly and provides a measure of the order parameter versus temperature, which is proportional to the O3 bond angle; in the paraelectric pseudohexagonal orthorhombic phase, the oxygen ion angle around the $Ca^{++}$ ion is ca. 130 degrees..

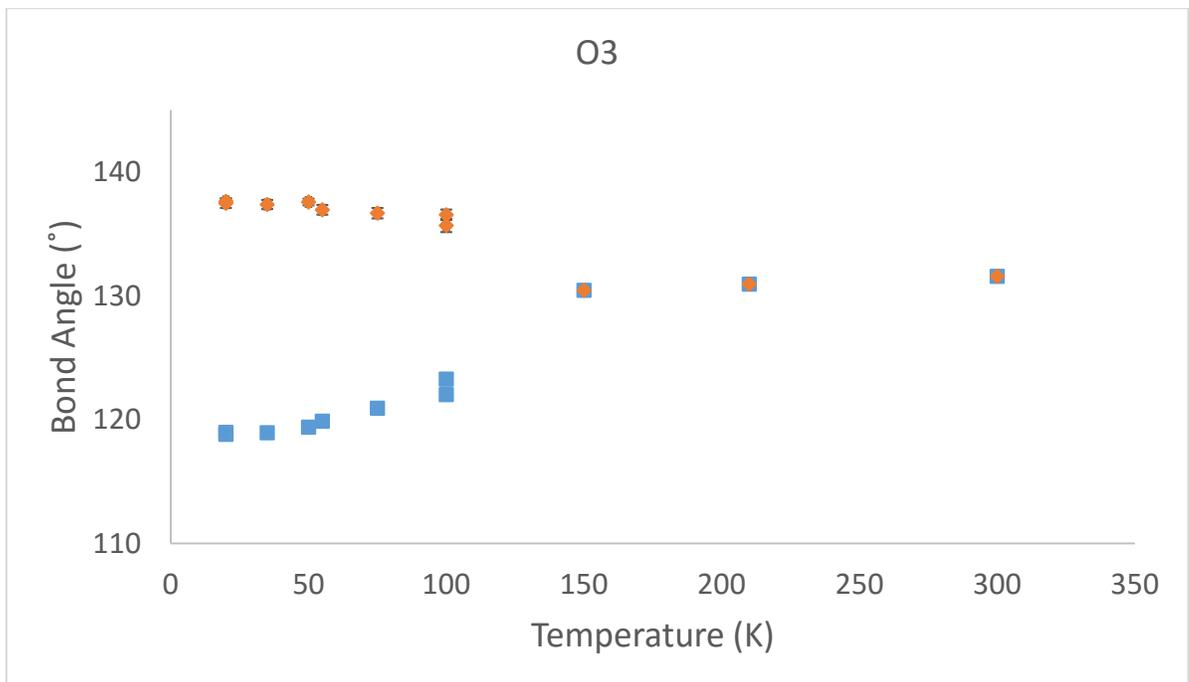

Fig.8. Bond angle of oxygen-3 ion versus temperature, scaling as the order parameter.

VI.    Relation to QCP (Quantum Critical Point)

Recently we showed[7] that despite being uniaxial, TSCC exhibits a critical exponent γ for its isothermal susceptibility (dielectric constant) of γ = 2, rather than the predicted 3. This was inferred to arise from the ultra-weak polarization in TSCC (0.33 µC/cm$^2$ at 77 K), which makes the system behave as if it were only negligibly ferroelectric at low T, similar to pseudo-cubic $SrTiO_3$ or $KTaO_3$. The present work supports that view and in fact shows that P(V=0) is zero below 64 K, due to very large bias voltages (built-in potential); hence, the dielectric response at small voltages is that of a non-polar crystal. This is a stronger argument than in Ref. 7. To restate this conclusion: No measurements near V=0 voltage reveal ferroelectricity; only at high fields is switching measured.

VII.    **S**ummary

We present polarization hysteresis loops in TSCC below its known phase transition at 64 K (pure TSCC; 76 K in TSCC with a few percent Br). These data show that the low-temperature phase cannot be $C_{2h}$ point group, as proposed by Roth et al.; it cannot be centric at all. This inference is supported by Raman data. The latter show some similarities with the high-pressure, room-temperature phase, but the structures are not identical; and we emphasize that Schmidt's extrapolated phase boundary does not predict that the high-pressure phase can be reached at 1 bar even at T=0, and Fujimoto et al. show that this phase is not polar. The "phase" below 64 K appears to be ferrielectric, in the Lieb-Mattis sense of ferrimagnetism (different numbers of spins of the same magnitude in antiparallel arrays).[25] The low-temperature structure is complicated by a probable additional transition near 48 K. However the X-ray and neutron structural study reveals no distortion to crystal class lower than orthorhombic.

A semantic question arises as to how best to describe "ferroelectrics" for which the bias field $E_b$ >> the coercive field $E_c$: these crystals have no ferroelectric properties for voltages near zero. These are well known as arising from defects and substitutional ions in materials such as TGS (tri-glycine sulphate), but they may also arise from a ferroelectric structure with several inequivalent polar sites (e.g., sarcosine S1 and S2 sites in the present work). In the latter case they may have two local polarizations, one of which switches and the other of which does not; they are reminiscent of Lieb-Mattis models of ferrimagnetism.

**References:**

[1]G.E. Feldkamp, J. F. Scott, W. Windsch, Light-Scattering Study of Phase Transitions in Ferroelectric tris-Calcium Chloride and its Brominated Isomorphs, *Ferroelectrics* **39**, 1163 (1981).

[2]S. P. P. Jones, D. M. Evans, M. A. Carpenter, S. A. T. Redfern, J. F. Scott, U. Straube, V. H. Schmidt, Phase diagram and phase transitions in tris-sarcosine calcium chloride and its brominated isomorphs, *Phys. Rev. B* **83**, 094102 (2010).

[3]J. Bornarel, V. H. Schmidt, Determination of Landau free-energy parameters by dielectric measurements in the ferroelectric TSCC, *J. Phys. C: Sol. St.* **14**, 2017 (1981).

[4](a) S. Fujimoto, N. Yasuda, H. Kashiki, K. Takagi, M. Fujimoto, Dielectric properties of solid solutions of tris-sarcosine calcium chloride TSCC and bromide TSCB, *Ferroelectrics* 39, 1139 (1981); (b) S. Fujimoto, N. Yasuda, A. Kawamura, T. Hachiga, Temperature and pressure-dependence of dielectric-properties of iodinated tris-sarcosine calcium-chloride, *J. Phys. D* **17**, 1019 (1984).

[5]R. Mackeviciute, M. Ivanov, J. Banys, N. Novak, Z. Kutnjak, M. Wencka, J. F. Scott, The Perfect Soft-Mode: Giant Phonon Instability in a Ferroelectric, *J. Phys. Cond. Mat* .**25**, 212201 (2013).

[6]J. C. Lashley, J. H. D. Munns, M. Echizen, M. N. Ali, S. E. Rowley, J. F. Scott. Phase transitions in the brominated ferroelectric tris-sarcosine calcium chloride (TSCC), *Adv. Mater.* **26**, 3860 (2014); S. E. Rowley, M. Hadjimichael, M. Ali, Y. C. Durmaz, J. Lashley, R. J. Cava,' J. F. Scott, Quantum criticality in a uniaxial organic ferroelectric, *J. Phys. Cond. Mat.* **27**, 395901 (2015).

[8]R. Roth, G. Schaack, H. D. Hochheimer, Raman study of tris-sarcosine calcium chloride in

the antiferroelectric phase, *Sol. St. Commun.* **55**, 121 (1985).

[9]V. H. Schmidt, Dielectric observation of a probably anti-ferroelectric high-pressure phase in the ferroelectric tris-sarcosine calcium chloride TSCC, *Sol St Commun* **35**, 649-652 (1980).

[10]J. Bornarel, V. H. Schmidt, Determination of Landau free-energy parameters by dielectric measurements in the ferroelectric TSCC, *J. Phys. C : Sol. St.* **14**, 2017 (1981).

[11]L. E. Cross, A thermodynamic treatment of ferroelectrics and antiferroelectrics in pseudo-cubic dielectrics, *Phil. Mag.* **1**, 76 (1956).

[12]C. F. Pulvari, Ferrielectricity, *Phys. Rev.* **120**, 1670 (1960).

[13]Z. Y. Zhao, M. F. Lin, X. Li, L. Lin, Z. B. Yan, S. Dong, J-M Liu, Experimental observation of ferrielectricity in multiferroic $DyMn_2O_5$, *Sci. Rpts.* **4**, 3984 (2014).

[14]N. Hur, S. Park, P. A. Sharma, S. Guha, and S. W. Cheong, Colossal magnetodielectric effects in $DyMn_2O_5$, *Phys. Rev. Lett.* **93**, 107207 (2004).

[15]A. Dziaugys, J. Banys, J. Macutkevic, R. Sobiestianskas, and Y. Vysochanskii, Dipolar glass phase in ferrielectrics: $CuInP_2S_6$ and $Ag_{0.1}Cu_{0.9}InP_2S_6$ crystals, *Phys. Stat. Sol.* **207**, 1960 (2010).

[16]S. Jaradat, P. D. Brimicombe, N. W. Roberts, C. Southern, H. F. Gleeson, Asymmetric switching in a ferrielectric liquid crystal device, *Appl. Phys. Lett.* **93**, 153506 (2008).

[17]S. Jaradat, P. D. Brimicombe, C. Southern, S. D. Siemianowski, E. DiMasi, R. Pindak, H. F. Gleeson, Stable field-induced ferrielectric liquid crystal phases in devices, *Appl. Phys. Lett.* **94**, 153507 (2009)

[18]G. Gou and J. M. Rondinelli, Piezoelectricity across a Strain-Induced Isosymmetric Ferri-to-Ferroelectric Transition, *Adv. Mater. Interfaces* **1**, 1400042 (2014).

[19]H. Haga, A. Onodera, H. Yamashita, Y. Shiozaki, New phase transition in TSCC at low temperatures, *J. Phys. Soc. Jpn.* **62,** 1857 (1993); 1857.

[20]K. Lee, M. Lee, K. S. Lee, A. R. Lim, $H^1$ NMR study of the phase transitions of tris-sarcosine calcium chloride single crystals at low temperature, *J. Phys. Chem. Sol.* **66**, 1739 (2005).

[21]C. A. Randall, private communication (2015).

[22]H. Shimizu, H. Guo, S. E. Reyes-Lillo, Y. Mizuno, K. M. Rabe, C. A. Randall CA, Lead-free antiferroelectric xCaZrO$_3$-(1-x)NaNbO$_3$ system (0,x,0.10), *Dalton Trans.* **44**, 10763 (2015).

[23]H. Guo, H. Shimizu, Y. Mizuno, C. A. Randall, Strategy for stabilization of the antiferroelectric phase Pbma over the metastable ferroelectric phase P2$_1$ma in lead-free (1-x)[NaNbO$_3$]-x[SrZrO$_3$], *J. Appl. Phys.* **117**, 2144103 (2015).

[24]K. Dimmler, M. Parris, D. Butler, S. Eaton, B. Pouligny, J. F. Scott, and Y. Ishibashi, Switching Kinetics in KNO$_3$ Ferroelectric Thin-film Memories, *J. Appl. Phys.* **61**, 5467 (1987).

[25]E. Lieb and D. Mattis, Ordering energy levels of interacting spin systems, *J. Math. Phys.* **3**, 749 (1962).